\documentclass[twocolumn,tightenlines,amsmath,amssymb]{revtex4-2}

\usepackage{bm}% bold math
\usepackage[colorlinks,anchorcolor=black,citecolor=violet,linkcolor=blue,urlcolor=brown]{hyperref}
\usepackage[mathscr]{euscript}
\usepackage{graphicx}% Include figure files
\usepackage{multirow,slashed}
\usepackage{xcolor}

\begin{document}

\parskip=3pt

%\preprint{}

\title{Pursuit of \boldmath$CP$ violation in hyperon decays at $e^+e^-$ colliders}

\author{Xiao-Gang He\vspace{-2ex}}\email{hexg@sjtu.edu.cn}
\affiliation{Tsung-Dao Lee Institute, KLPAC and SKLPPC laboratories, School of Physics and Astronomy,
Shanghai Jiao Tong University, Shanghai 200240, China \smallskip \\
Department of Physics, National Taiwan University, Taipei 10617, Taiwan\vspace{-1ex}}

\author{Jusak Tandean\vspace{-2ex}}\email{jtandean@yahoo.com}
\affiliation{Gayungsari III/20, Surabaya 60235, Indonesia\vspace{-1ex}}

\author{German Valencia\vspace{-2ex}}\email{german.valencia@monash.edu}
\affiliation{School of Physics and Astronomy, Monash University, Melbourne VIC-3800, Australia}

%\date{\vspace*{3ex}}

\maketitle

The phenomenon of $CP$ violation---the breaking of the combination of charge-conjugation symmetry ($C$) and parity symmetry ($P$)---is one of the least understood aspects of high-energy physics.
The first evidence for $CP$ violation was reported in 1964 by a group led by Cronin and Fitch~\cite{Christenson:1964fg} after studying neutral-kaon decays.
It was later recognized in 1967 by Sakharov~\cite{Sakharov:1967dj} that $CP$ violation is an important ingredient for resolving the mystery of why our Universe is dominated by matter over antimatter if they were created in equal quantities by the Big Bang.
The quest for understanding $CP$ violation has since drawn a great deal of attention from both the particle physics and cosmology communities.
In 1973 Kobayashi and Maskawa~\cite{Kobayashi:1973fv} showed that $CP$ violation would emerge naturally in the presence of three families of quarks, establishing its mechanism in the standard model (SM) of particle physics.

After the aforesaid historical discovery of $CP$ violation, numerous different searches for it were carried out.
Some of them yielded positive results, empirically confirming its existence not only in the kaon sector but also in the beauty-meson and charmed-meson systems, and all of the observations to date are compatible with the SM expectations~\cite{Zyla:2020zbs}.
Despite its successes, however, it is widely believed that the SM falls short of being the complete theory of fundamental interactions.
One of the reasons is that the amount of $CP$ violation that the SM could generate is thought to be inadequate if it is to be solely responsible for the matter-antimatter imbalance of the Universe~\cite{Canetti:2012zc}.
This suggests the necessity of continuing to hunt for signals of $CP$ violation in many other processes, the outcomes of which could further check its SM explanation and, moreover, might help pin down potential sources of $CP$ violation from beyond the SM.

Outside the meson sectors mentioned above, baryon systems have not yet supplied firm indications of $CP$ violation, but there are ongoing and forthcoming efforts to change the situation, a few of which will be discussed shortly.
Having odd half-integer spins, baryons can offer environments where additional tests of $CP$-violation may be available besides the traditional particle-antiparticle decay-rate differences and mixing-related observables.

Hereafter we focus on the lightest spin-1/2 hyperons (strange-flavored baryons), specifically $\Lambda$, $\Sigma$, and $\Xi$.
Their nonleptonic two-body decays could provide several observables potentially sensitive to $CP$-violating effects, as pointed out decades ago~\cite{Okubo:1958zza,Pais:1959zza,Brown:1983wd,Chau:1983ei,Donoghue:1985ww,Donoghue:1986hh,Donoghue:1986nn,He:1991pf,CP-HyperonStudyGroup:1992qvi}.
The $\Omega^-$ hyperon, which has spin 3/2, could play an analogous role~\cite{Tandean:1998wr,Tandean:2004mv}.

Early experimental endeavors to realize this were performed from the 1980s through 2010 by a number of groups~\cite{Zyla:2020zbs,Barnes:1987vc,DM2:1988ppi,HyperCP:2004zvh,HyperCP:2006ktj,BES:2009zvb} investigating the decays of $\Lambda$ and $\Xi^-$ as well as $\Omega^-$.
Most recently there has been a renewal of similar attempts at the Beijing Electron-Positron Collider II by the BESIII Collaboration~\cite{Bigi:2017eni,BESIII:2018cnd,BESIII:2020fqg,BESIII:2021ypr,BESIII:2022qax,BESIII:2022lsz}.
In its clean high-intensity environment, a charmonium resonance can form in $e^+e^-$ annihilation and subsequently convert into a quantum-entangled pair of hyperon and antihyperon which decay nonleptonically, allowing for the simultaneous determination of various parameters pertaining to $CP$ violation in the decays.
Its predecessors, DM2~\cite{DM2:1988ppi} and BES~\cite{BES:2009zvb}, utilized $e^+e^-$ collisions as well but had accumulated datasets which were much smaller than that of~BESIII.
Although all of the foregoing searches have so far come up empty, there remains open a large window into possible new physics beyond the SM to be explored by BESIII in its current program.
Further in the future, the quests are anticipated to be taken up by the proposed super tau-charm factories~\cite{Zhou:2021rgi,Barnyakov:2020vob}, with total luminosities increased by two orders of magnitude, which might enable them to attain SM-level sensitivity.

It is worth remarking that one could also seek $CP$-violation signs in the concurrent decays of a hyperon and antihyperon produced pairwise in proton-antiproton scattering~\cite{Donoghue:1986nn,CP-HyperonStudyGroup:1992qvi}.
This was tried in the past~\cite{Zyla:2020zbs,Barnes:1987vc}, but with limited statistics, and will be undertaken again by the upcoming PANDA experiment in Germany~\cite{PANDA:2020zwv}.

In the main processes of interest here, a spin-1/2 hyperon, ${\cal B}_i$, turns into another spin-1/2 baryon, ${\cal B}_f$, and a pion, $\pi$, causing the strangeness quantum number to change by one unit.
This is described by an amplitude expressible as ${\cal M} = \chi_f^\dagger \big(S+\bm{\sigma} \cdot \hat{\textbf{\textit p}}_{\!f} P\big)\chi_i^{}$, where $S$ and $P$ are generally complex constants corresponding to parity-violating S-wave and parity-conserving P-wave transitions, respectively, $\bm{\sigma}$ and $\chi_f^{},\chi_i^{}$ represent Pauli matrices and spinors, and $\hat{\textbf{\textit p}}_{\!f}$ is the momentum unit-vector of ${\cal B}_f$.
In the rest frame of ${\cal B}_i$, if it has a polarization $\textbf{\textit P}_{\!i}$, the angular distribution of ${\cal B}_f$ is given by
\begin{align}
\frac{d\Gamma_{{\cal B}_i\to {\cal B}_f\pi}}{d\Omega_f} & = \frac{\Gamma_{{\cal B}_i\to {\cal B}_f\pi}}{4\pi}\big(1+\alpha\, \textbf{\textit P}\!_i\cdot\hat{\textbf{\textit p}}_{\!f}\big) &
\end{align}
and its polarization
\begin{equation}
\textbf{\textit P}\!_f = \frac{\big(\alpha + \textbf{\textit P}_{\!i} \!\cdot\! \hat{\textbf{\textit p}}_{\!f}\big)  \hat{\textbf{\textit p}}_{\!f} + \beta \textbf{\textit P}\!_i \!\times\! \hat{\textbf{\textit p}}_{\!f} + \gamma\, \hat{\textbf{\textit p}}_{\!f} \!\times\! \big(\textbf{\textit P}\!_i \!\times\! \hat{\textbf{\textit p}}_{\!f}\big)}{1 + \alpha\, \textbf{\textit P}\!_i \cdot \hat{\textbf{\textit p}}_{\!f}} ,
\end{equation}
where $\Gamma_{{\cal B}_i\to {\cal B}_f\pi}$ denotes the partial rate, $\Omega_f$ is the solid angle of ${\cal B}_f$, and the parameters $\alpha$, $\beta$, and $\gamma$ depend on $S$ and $P$ according to
\begin{equation}
\alpha = \frac{2\,{\rm Re}(S^*P)}{|S|^2+|P|^2} \,, ~ \beta = \frac{2\,{\rm Im}(S^*P)}{|S|^2+|P|^2} \,, ~ \gamma = \frac{|S|^2-|P|^2}{|S|^2+|P|^2}
\end{equation}
and satisfy $\alpha^2+\beta^2+\gamma^2=1$.
Another parameter $\phi$, which is linked to the second and third by $\beta=\sqrt{1-\alpha^2} \sin\phi$ and $\gamma=\sqrt{1-\alpha^2} \cos\phi$, and $\alpha$ are the most closely connected to experiment and essentially uncorrelated~\cite{Zyla:2020zbs}.
For the antiparticle process $\overline{{\cal B}_i}\to\overline{{\cal B}_f}\overline\pi$, the amplitude
has the form \,$\overline{\cal M} = \chi_f^\dagger \big(\overline S+\bm{\sigma} \cdot \hat{\textbf{\textit p}}_{\!f} \overline P\big)\chi_i^{}$ and the partial rate, $\overline\alpha$, $\overline\beta$, etc. are related to $\overline S$ and $\overline P$ in the same way as $\Gamma_{{\cal B}_i\to {\cal B}_f\pi}$, $\alpha$, $\beta$, etc. to $S$ and $P$.

Before addressing the $CP$ tests of interest, we write in the customary parameterization
\begin{align} \label{s,p}
S & = \raisebox{2pt}{\footnotesize$\displaystyle\sum_r$}\, S_r^{}\, e^{i(\delta_S^r+\xi_S^r)} \,, & \overline S & = -\raisebox{2pt}{\footnotesize$\displaystyle\sum_r$}\, S_r^{}\, e^{i(\delta_S^r-\xi_S^r)} \,, \nonumber \\
P & = \raisebox{2pt}{\footnotesize$\displaystyle\sum_r$}\, P_r^{}\, e^{i(\delta_P^r+\xi_P^r)} \,, & \overline P & = \raisebox{2pt}{\footnotesize$\displaystyle\sum_r$}\, P_r^{}\, e^{i(\delta_P^r-\xi_P^r)} \,,
\end{align}
where the index $r$ in the sums runs over all possible amplitudes for different final-isospin states and changes $\Delta I$ in isospin, $S_r$ and $P_r$ are real numbers, $\delta_S^r$ and $\delta_P^r$ designate the $CP$-conserving phase-shifts arising from strong rescattering of the final states, and $\xi_S^r$ and $\xi_P^r$ stand for weak phases encoding the $CP$ violation.
In eq.\,(\ref{s,p}) the extra minus sign of $\overline S$ relative to $\overline P$ and the flipped signs of $\xi_S^r$ and $\xi_P^r$ in them compared to $S$ and $P$ are attributable to the assumed invariance of the interaction Hamiltonian under the combined $CP$ and time-reversal transformation and to the application of unitarity.
If $CP$ symmetry holds, $\xi_S^r=\xi_P^r=0$ in eq.\,(\ref{s,p}), which implies that $\alpha=-\overline\alpha$ and $\beta=-\overline\beta$.
It follows that
\begin{align} \label{A,B}
A_{CP}^{} & = \frac{\alpha+\overline\alpha}{\alpha-\overline\alpha} \,, & B_{CP}^{} & = \frac{\beta+\overline\beta}{\alpha-\overline\alpha} &
\end{align}
are tests of $CP$ violation.
Each of them would probe, at the same time, the parity-even (P-wave) and parity-odd (S-wave) parts of the responsible interactions.
Therefore, its measurements could supply valuable information which is complementary to that gained from experimental studies on the so-called indirect and direct types of $CP$ violation in the kaon system, which originate from the parity-even and -odd components, respectively, of the underlying physics.

For ${\cal B}_i=\Lambda$ or $\Xi$, it is known from data that the term corresponding to $\Delta I = 1/2$ dominates each of $S$, $P$, $\overline S$, and $\overline P$ in eq.\,(\ref{s,p}).
Accordingly, the leading expression for $A_{CP}$ is~\cite{Donoghue:1986hh}
\begin{align} \label{acp}
A_{CP}^{} & = -\tan(\delta_P-\delta_S)\,\tan(\xi_P-\xi_S) \,,
\end{align}
where the $r$ superscript has been dropped for visual simplicity.
The majority of past searches for hyperon $CP$-violation targeted this asymmetry~\cite{Zyla:2020zbs}.

It is clear from eq.\,(\ref{acp}) that to make $A_{CP}$ nonzero requires both $\delta_P\neq\delta_S$ and $\xi_P\neq\xi_S$ and that a small phase-shift difference, $\delta_P-\delta_S$, will bring about a suppressive factor in $A_{CP}$ which reduces its sensitivity to $CP$-violating effects.
Moreover, extracting a precise value of $\xi_P-\xi_S$ from it will be unattainable if the phase shifts are poorly known.
An observable that is not subject to these issues is $B_{CP}$ in eq.\,(\ref{A,B}) which at leading order becomes~\cite{Donoghue:1985ww,CP-HyperonStudyGroup:1992qvi}
\begin{align} \label{bcp}
B_{CP}^{} & = \tan(\xi_P-\xi_S) \,,
\end{align}
independent of $\delta_{P,S}$.
As a consequence, $B_{CP}$ would exceed $A_{CP}$ in size by up to an order of magnitude~\cite{Donoghue:1985ww}.
There is, however, a limitation to pursuing $B_{CP}$, namely that $\beta$ is more challenging to measure than $\alpha$ if one relies on a single-step transition ${\cal B}_i\to {\cal B}_f\pi$, as this would have to involve a dedicated detector for the polarization of ${\cal B}_f$.
We note that the usual partial-rate asymmetry is even harder to reach than $A_{CP}$ due to additional suppression by small $\Delta I=3/2$ contributions.

As it turns out, it is possible to derive simultaneously not only $\alpha$ and $\phi$ of ${\cal B}_i\to {\cal B}_f\pi$ but also $\overline\alpha$ and $\overline\phi$ of
$\overline{{\cal B}_i}\to\overline{{\cal B}_f}\,\overline\pi$ if ${\cal B}_i$ and $\overline{{\cal B}_i}$ are pair-produced in the decay of a charmonium ({\it e.g.}, the $J/\psi$ meson) created by $e^+e^-$ annihilation and each undergo a nonleptonic two-step weak decay.
After translating the findings into $\beta$ and $\overline\beta$, the value of $B_{CP}$ can be obtained.
This has recently been achieved by BESIII~\cite{BESIII:2021ypr}, upon implementing a recently proposed method~\cite{Perotti:2018wxm,Adlarson:2019jtw} to the reaction $e^-e^+\to J/\psi\to\Xi^-\overline\Xi{}^+$ and the ensuing sequential transitions $\Xi^-\to\Lambda\pi^-\to p\pi^-\pi^-$ and
$\overline\Xi{}^+\to\overline\Lambda\pi^+\to\overline p\pi^+\pi^+$.
The acquired data led to the determination of $\alpha_\Xi$ and $\phi_\Xi$ for $\Xi^-\to\Lambda\pi^-$ and $\overline\alpha{}_\Xi$ and $\overline\phi{}_\Xi$ for $\overline\Xi{}^+\to\overline\Lambda\pi^+$, as well as $\alpha_\Lambda$ for $\Lambda\to p\pi^-$ and $\overline\alpha{}_\Lambda$ for $\overline\Lambda\to\overline p\pi^+$.
This allowed BESIII to report the $CP$ asymmetries~\cite{BESIII:2021ypr}
\begin{align} \label{bes3}
A_{CP}^\Xi & = (6\pm 13\pm 6)\times10^{-3} \,,
\nonumber \\
B_{CP}^\Xi & \simeq\, \xi_P^\Xi-\xi_S^\Xi = (1.2\pm 3.4\pm 0.8)\times10^{-2} \,, ~~~
\nonumber \\
A_{CP}^\Lambda & = (-4\pm 12\pm 9)\times10^{-3}
\end{align}
where the first and second uncertainties are statistical and systematic, respectively, and the superscript $\Xi$ ($\Lambda$) refers to the \,$\Xi\to\Lambda\pi$ ($\Lambda\to p\pi$) case.
The $CP$-conserving phase-shifts $(\delta_P-\delta_S)$ in the $\Xi$ modes were also derived, but again with sizable uncertainties~\cite{BESIII:2021ypr}.
It is worth mentioning that the $A_{CP}^\Xi$ and $B_{CP}^\Xi$ measurements had never been done before and that the $A_{CP}^\Lambda$ result agrees, and is compatible in precision, with the best earlier measurement~\cite{BESIII:2018cnd}.
Very recently BESIII~\cite{BESIII:2022qax} has announced the improved value $A_{CP}^\Lambda=(-2.5\pm 4.6\pm 1.1)\times10^{-3}$, which is the most precise to date, after directly examining \,$e^-e^+\to J/\psi\to\Lambda\overline\Lambda\to p\pi^-\overline p\pi^+$.

It is of interest to compare these BESIII findings to the expectations within and beyond the SM.
Currently the theoretical estimates suffer from considerable uncertainties and therefore can yield only ranges of values~\cite{Tandean:2002vy,Tandean:2003fr}.
For our quantities of concern, the SM predicts
\begin{align} \label{sm}
0.5\times10^{-5} & \le \big(A_{CP}^\Xi\big){}_{\rm SM}^{}\le 6\times10^{-5} \,,
\nonumber \\
-3.8\times10^{-4} & \le \big(\xi_P^\Xi-\xi_S^\Xi\big){}_{\rm SM}^{}\le -0.3\times10^{-4} \,, ~~~
\nonumber \\
-3\times10^{-5} & \le \big(A_{CP}^\Lambda\big){}_{\rm SM}^{}\le 3\times10^{-5} \,,
\end{align}
as updated in ref.\,\cite{Salone:2022lpt} from the analysis of ref.\,\cite{Tandean:2002vy}.
These are below their empirical counterparts given in the previous paragraph by two orders of magnitude or more and might be out of BESIII scope ultimately.
Nevertheless, the predictions in eq.\,(\ref{sm}) could be within reach of the proposed super tau-charm factories~\cite{Zhou:2021rgi,Barnyakov:2020vob} because their total luminosities are planned to surpass that of BESIII by two orders of magnitude, implying better sensitivity by at least ten times, and further improvement can be gained if polarized-electron beams are utilized~\cite{Salone:2022lpt}.

\begin{figure}[b]
\includegraphics[width=77mm]{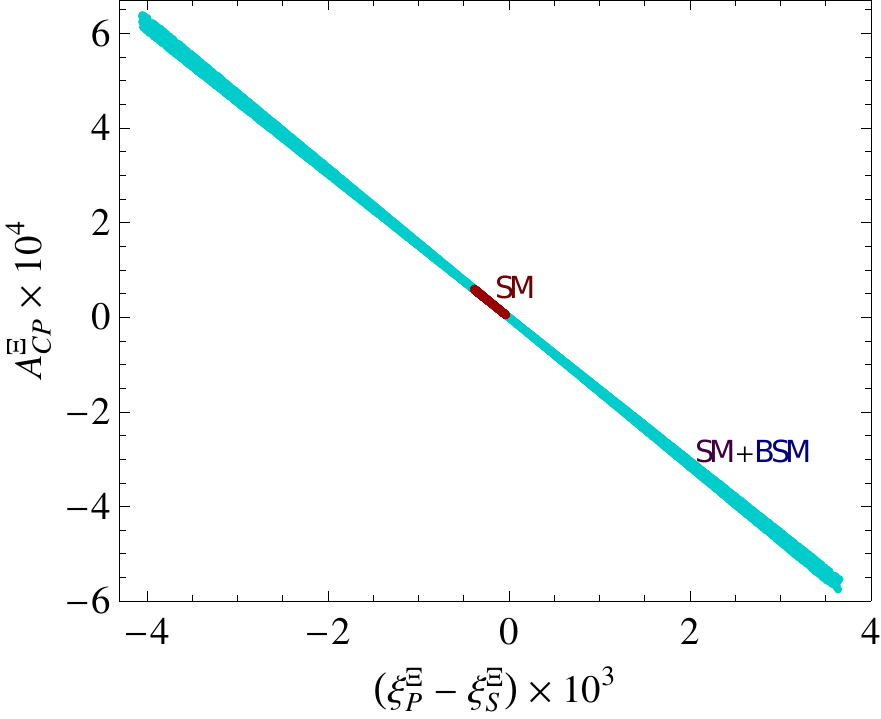} \vspace{-7pt}
\caption{Correlation between theoretical $A_{CP}^\Xi$ and $\xi_P^\Xi-\xi_S^\Xi$.
The (darkly shaded) red patch comes from the SM alone and lies within the (lightly shaded) cyan region which includes the BSM contribution described in the text.\label{A-xi}}
\vspace{7ex}

\includegraphics[width=77mm]{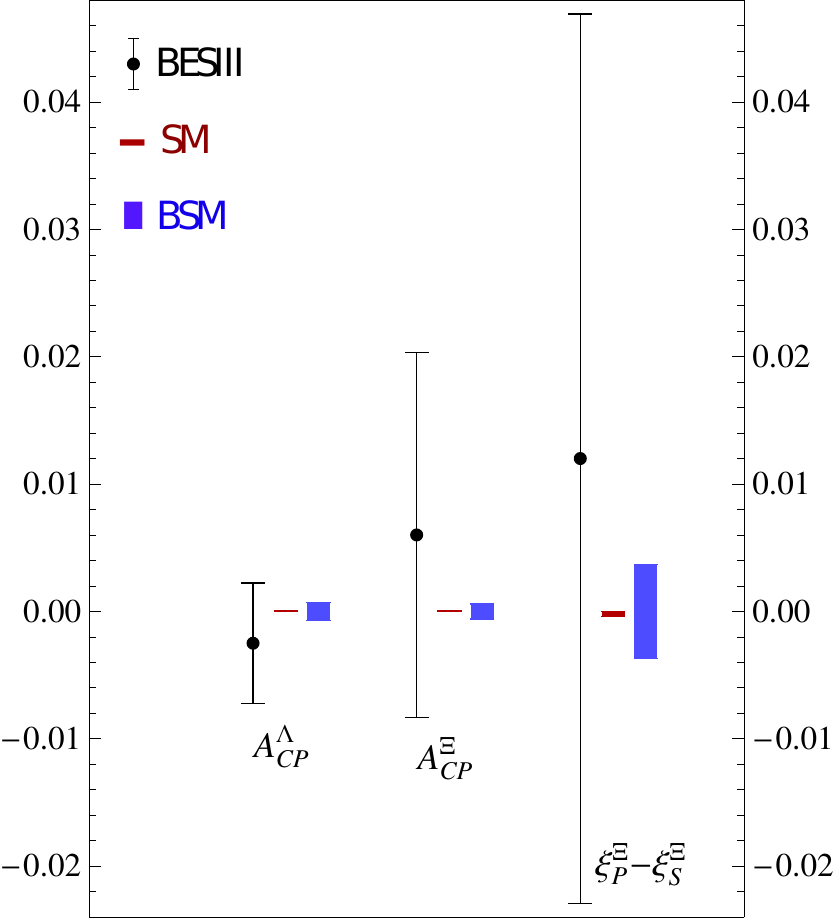}
\caption{The latest BESIII results for $A_{CP}^\Xi$ and $\xi_P^\Xi-\xi_S^\Xi$ from ref.\,\cite{BESIII:2021ypr} and $A_{CP}^\Lambda$ from ref.\,\cite{BESIII:2022qax} (the statistical and systematic uncertainties combined in quadrature) compared to their respective theoretical predictions.\label{exp-th} }
\end{figure}

In the presence of physics beyond the SM (BSM), there could be new sources of $CP$ violation
which enhance $A_{CP}$ and $\xi_P-\xi_S$ in size with respect to the SM expectations.
This possibility can be realized with, for instance, the so-called chromomagnetic-penguin
interactions~\cite{He:1995na,He:1999bv}.
Since their parity-even and -odd components also affect, respectively, the indirect and direct
$CP$-violation in the kaon system, restrictions from the relevant kaon data apply.
Taking the latter into account and employing the work of ref.\,\cite{Tandean:2003fr}, with parameter inputs from ref.\,\cite{Salone:2022lpt}, we arrive at
\,$|A_{CP}^\Xi|_{\rm BSM}^{}<5.9\times10^{-4}$,\,
$|\xi_P^\Xi-\xi_S^\Xi|_{\rm BSM}^{}<3.7\times10^{-3}$, and
\,$|A_{CP}^\Lambda|_{\rm BSM}^{}<7.0\times10^{-4}$ for the chromomagnetic-penguin contributions.
These numbers are higher than the corresponding ones in eq.\,(\ref{sm}) from the SM by about an order of magnitude.
The correlation between $\xi_P^\Xi-\xi_S^\Xi$ and $A_{CP}^\Xi$ in the two scenarios is displayed in fig.\,\ref{A-xi}.
On the other hand, the preceding BSM results are still lower than their experimental counterparts quoted earlier by roughly a factor of ten.
The situation is depicted in fig.\,\ref{exp-th}.
Notwithstanding, forthcoming findings from BESIII may yet be able to test these predictions and in the future the super tau-charm factories would certainly probe them more stringently.

In conclusion, hyperon nonleptonic decays potentially serve as a promising hunting ground for $CP$ violation in the strangeness-changing sector caused by new physics beyond the SM.
Given that they involve fermions, these processes could supply information about the underlying interactions which is different from that furnished by kaon transitions.
It is therefore exciting that there are now quests for hyperon $CP$-violation that are ongoing intensively by BESIII and also planned or proposed ones at other facilities.
It is hoped that fresh hyperon data will appear soon which reveal evidence for new physics or, if not, meaningfully constrain BSM contributions.

As final remarks, since charmed baryons are copiously produced at running $e^+e^-$ machines and the LHC and some of them can undergo a Cabibbo-favored nonleptonic or semileptonic decay into a charmless final-state containing a hyperon with a branching fraction at the percent level or bigger~\cite{Zyla:2020zbs}, it might additionally be feasible to look for hyperon $CP$-violation by means of such processes~\cite{Salone:2022lpt,Wang:2022tcm}.
The first attempt in this direction has very recently been conducted by the Belle Collaboration~\cite{Belle:2022uod} studying $\Lambda_c^+\to\Lambda\pi^+$, although the search came up negative.

%\acknowledgments

\bigskip

{\bf\small Acknowledgments}~ This work was supported in part by the Australian Government through the Australian Research Council. XGH was supported in part by the NSFC (Grant Nos. 11735010, 11975149, and 12090064), and also supported in part by the MOST (Grant No. MOST 109-2112-M-002-017-MY3).

\end{document}